\documentclass[envcountsame]{llncs}

\usepackage{graphicx} 
\usepackage{times}
\usepackage{amssymb,amsmath}
\usepackage{gastex}
\usepackage{epsfig}
\usepackage{liststyle}

\usepackage[ruled, vlined]{algorithm2e}
\dontprintsemicolon

\newcounter{compressEnum}

{}

\newtheorem{mydefi}{Definition}

        {\qed\par\smallskip\global\def\qed{\origQED\global\def\qed{}}}

\def\endof{%
  \leavevmode
  \parfillskip=0pt%
  \widowpenalty=10000%
  \displaywidowpenalty=10000%
  \finalhyphendemerits=0%
  \unskip\nobreak\null\hfil\penalty50\hskip2em\null\hfill%
}
\def\eodsymbol{\ensuremath\square}
\def\eopsymbol{\ensuremath\blacksquare}

\def\origEOD{\nobreak\leavevmode\endof\eodsymbol\par}
\def\EOD{\origEOD\global\def\EOD{}}
\def\origQED{{\nobreak\leavevmode\endof\eopsymbol\par\smallskip}}
\def\qed{\origQED\global\def\qed{}}

\newcommand{\li}{l_0}

\def\abs#1{\ensuremath{\lvert #1\rvert}}

\newcommand{\post}{\mathsf{post}}

\newcommand{\Outcome}{{\mathsf{Outcome}}}

\newcommand{\trans}{{\Delta}}

\newcommand{\nat}{\mathbb N}

\newcommand{\real}{{\mathbb R}}

\newcommand{\arup}[1]{\lceil \mathrel {#1} \rceil}

\newcommand{\bigarupL}{\big\lceil}

\newcommand{\bigarupR}{\big\rceil}

\renewcommand{\l}{{\ell}}

\newcommand{\true}{{\sf true}}

\newcommand{\tuple}[1]{( #1 )}

\newcommand{\CPre}{\mathsf{CPre}}
\newcommand{\CP}{\mathsf{CP}}

\newcommand{\Obs}{\Gamma}

\newcommand{\obs}{\mathsf{obs}}

\newcommand{\si}{s_0}
\newcommand{\straa}{\alpha}

\def\rank{{\sf rank}}

\RequirePackage{xspace}
\def\ie{i.e.,\xspace}

\newcounter{menum}

\listspacing{in}{\topsep= 4pt plus 1pt minus 2pt\parsep=  2pt plus 1pt   minus 0.5pt\itemsep= 2pt plus 1.5pt minus 0.5pt}
\listspacing{is}{\topsep= 2pt plus 1pt minus 1pt\parsep=  1pt plus 0.5pt minus 0.5pt\itemsep= 1pt plus 0.5pt minus 0.5pt}
\listspacing{if}{\topsep= 2pt plus 1pt minus 0.5pt\parsep=1pt plus 0.5pt minus 0.5pt\itemsep= 1pt plus 0.5pt minus 0.5pt}
\listspacing{ii}{\topsep= 2pt plus 1pt minus 0.5pt\parsep=1pt plus 0.5pt minus 0.5pt\itemsep= 1pt plus 0.5pt minus 0.5pt}
\listspacing{iii}{\topsep=2pt plus 1pt minus 0.5pt\parsep=0pt\itemsep= 1pt plus 0.5pt minus 0.5pt}
\listspacing{iv}{\topsep= 2pt plus 1pt minus 0.5pt\parsep=0pt\itemsep= 1pt plus 0.5pt minus 0.5pt}
\listspacing{v}{\topsep=  2pt plus 1pt minus 0.5pt\parsep=0pt\itemsep= 1pt plus 0.5pt minus 0.5pt}
\listspacing{vi}{\topsep= 2pt plus 1pt minus 0.5pt\parsep=0pt\itemsep= 1pt plus 0.5pt minus 0.5pt}

\renewcommand{\sigma}{a}

\makeatletter

\begingroup \catcode `|=0 \catcode `[= 1
\catcode`]=2 \catcode `\{=12 \catcode `\}=12
\catcode`\\=12 |gdef|@xcomment#1\end{comment}[|end[comment]]
|endgroup

\def\@comment{\let\do\@makeother \dospecials\catcode`\^^M=10\def\par{}}

\def\begincomment{\@comment\@xcomment}

\makeatother

\begin{document}

\title{{\bf Alpaga: A Tool for Solving Parity Games\\ with Imperfect Information
}}


\author{Dietmar Berwanger\inst{1} \and Krishnendu Chatterjee\inst{2} \and Martin De Wulf\inst{3} \and \\ Laurent Doyen\inst{4} \and 
Thomas A. Henzinger\inst{4}}

\institute{
LSV, ENS Cachan and CNRS, France\\
\and CE, University of California, Santa Cruz, U.S.A.\\
\and Universit\'e Libre de Bruxelles (ULB), Belgium
\and \'{E}cole Polytechnique F\'ed\'erale de Lausanne (EPFL), Switzerland\\
}

\maketitle
\begin{abstract}
Alpaga is a solver for two-player parity games with imperfect information.
Given the description of a game, it determines whether the first player can
ensure to win and, if so, it constructs a winning strategy.  
The tool provides a symbolic implementation of a recent algorithm 
based on antichains. 
\end{abstract}

\section{Introduction}

Alpaga is a tool for solving parity games with imperfect information. 
These are games played on a graph by two players; the first player has
imperfect information about the current state of the play.
We consider objectives over infinite paths 
specified by parity conditions that can express safety, reachability,
liveness, fairness, and most properties commonly used in verification. 
Given the description of a game, the tool determines whether the
first player has a winning strategy 
for the parity objective and, if this is the case, 
it constructs such a winning strategy.

The Alpaga implementation is based on a recent technique 
using \emph{antichains} for solving
games with imperfect information efficiently~\cite{CDHR07}, 
and for representing the strategies compactly~\cite{BCDHR08}. 
To the best of our knowledge, this is the first implementation of a tool 
for solving parity games with imperfect information.

In this paper, we outline the antichain technique which is based on
fixed-point computations using a compact representation of sets. Our 
algorithm essentially iterates a \emph{controllable predecessor} operator
that returns the states from which a player can force the play into a 
given target set in one round. For computing this operator, 
no polynomial algorithms is known. We propose a new symbolic 
implementation based on BDDs to avoid the naive 
enumerative procedure.

Imperfect-information games arise in several important applications related to
verification and synthesis of reactive systems.
The following are some key applications: 
(a) synthesis of controllers for plants with unobservable transitions;
(b) distributed synthesis of processes with private variables not visible to 
other processes;
(c) synthesis of robust controllers;
(d) synthesis of automata specifications where only observations of automata 
are visible, and 
(e) the decision and simulation problem of quantitative specification languages;
(f) model-checking secrecy and information flow.
We believe that the tool Alpaga will make imperfect information 
games a useful framework for designers in the above applications.
In the appendix, we present a concrete example of distributed-system synthesis.
Along the lines of~\cite{ChaHen07}, 
we consider the design of a mutual-exclusion protocol for two
processes. The tool Alpaga is able to synthesize a winning strategy
for a requirement of mutual exclusion and starvation freedom which
corresponds to Peterson's protocol. 

\section{Games and Algorithms}\label{sec-games-algo}

Let $\Sigma$  be a finite alphabet of actions and let $\Obs$
be a finite alphabet of observations.      
A~\emph{game structure with imperfect information} over $\Sigma$ and
$\Gamma$ is a tuple $G=\tuple{L,\li,\trans,\gamma}$, where 
\begin{itemize}
\item $L$ is a finite set of locations (or states), 
\item $\li \in L$ is the initial location,
\item $\trans \subseteq L \times \Sigma \times L$ is a set of labelled 
transitions such that 
for all $\l \in L$ and all $\sigma \in \Sigma$, there exists 
$\l' \in L$ such that $(\l, \sigma, \l') \in \trans$, \ie the
transition relation is total,
\item $\gamma: \Obs \to 2^L \setminus \emptyset$ is an observability function 
that maps each observation to a set of locations such that the set $\{\gamma(o) \mid  o \in \Obs \}$ 
partitions $L$. 
For each $\l \in L$, let $\obs(\l) = o$ be the unique observation such that $\l \in \gamma(o)$. 
\end{itemize}

The game on $G$ is played in rounds. 
Initially, a token is placed in location~$\li$. 
In every round, Player~$1$ first chooses an action $\sigma \in \Sigma$, and 
then Player~$2$ moves the token to an $\sigma$-successor~$\l'$ of the current location~$\l$,
\ie such that $(\l, a, \l') \in \trans$. Player~$1$ does not see the
current location $\l$ of the token, but
only the observation $\obs(\l)$ associated to it.
%
A \emph{strategy} for Player~$1$ in~$G$ is a function 
$\straa:\Obs^+ \to \Sigma$. 
The set of possible \emph{outcomes} 
of $\straa$ in $G$ is the set $\Outcome(G,\straa)$ of sequences
$\pi=\l_1 \l_2 \ldots$ such that $\l_1 = \li$ and 
$(\l_i, \straa(\obs(\l_1\dots\l_i)), \l_{i+1}) \in \trans$ for all $i \geq 1$.
A \emph{visible parity condition} on $G$ is defined by a function 
$p:\Obs \to \nat$ that maps
each observation to a non-negative integer priority. 
We say that a strategy~$\straa$ for Player~$1$ is \emph{winning} if 
for all $\pi \in \Outcome(G,\straa)$, the least priority that appears infinitely 
often in~$\pi$ is even.

To decide whether Player~$1$ is winning in a game~$G$, 
the basic approach consists in tracing 
the \emph{knowledge} of Player~$1$, 
represented a set of locations called a \emph{cell}. 
The initial knowledge is the cell $\si = \{\li\}$. 
After each round, the knowledge $s$ of Player~$1$ is updated
according to the action $\sigma$ she played and the observation $o$ she receives,
to $s' =\post_\sigma(s) \cap \gamma(o)$ where $\post_{\sigma}(s) = 
\{\l' \in L \mid \exists \l \in s: (\l,a,\l') \in \trans \}$.


\paragraph{Antichain algorithm.}
The antichain algorithm is based on the \emph{controllable predecessor} operator
$\CPre: 2^S \to 2^S$ which, given a set of cells $q$, computes the set of
cells $q'$ from which Player~$1$ can force the game into a cell of $q$
in one round: 
\begin{equation}
\CPre(q)= \{ s \subseteq L \mid \exists \sigma \in \Sigma \cdot
\forall o \in \Obs: \post_\sigma(s) \cap \gamma(o) \in q \}.
\end{equation}
The key of the algorithm relies on the fact that $\CPre(\cdot)$ preserves
downward-closedness. A set $q$ of cells is \emph{downward-closed} if, for all
$s \in q$, every subset $s' \subseteq s$ is also in~$q$. 
Downward-closed sets~$q$ can be 
represented succinctly by their maximal elements 
$r = \arup{q} = \{s \in q \mid \forall s' \in q: s \not\subset s'\}$,
which form an \emph{antichain}.
With this representation, the controllable predecessor 
operator is defined by
\begin{equation}
  \CPre(r) = \bigarupL \{s \subseteq L \mid \exists \sigma \in \Sigma \cdot \forall o \in \Obs \cdot \exists s' \in r:
  \post_{\sigma}(s) \cap \gamma(o) \subseteq s'\} \bigarupR. \label{eq:cpre}
\end{equation}

\paragraph{Strategy construction.} 
The implementation of the strategy construction is based 
on~\cite{BCDHR08}.
The algorithm of~\cite{BCDHR08} employs antichains to compute winning
strategies for imperfect-information parity games in an efficient and
compact way: the procedure is similar to the classical algorithm of
McNaughton~\cite{McNaughton93} and  
Zielonka~\cite{Zielonka98} for perfect-information parity games, 
but, to preserve downwards closure, 
it avoids the complementation operation of the classical
algorithms by recurring into subgames with an objective obtained as a  
boolean combination of reachability, safety, and reduced parity
objectives.



\paragraph{Strategy simplification.} 
A strategy in a game with imperfect information can be represented by a 
set $\Pi=\{(s_1,\rank_1,\sigma_1),\dots,(s_n,\rank_n,\sigma_n)\}$
%
of triples $(s_i,\rank_i,\sigma_i) \in 2^L \times \nat \times \Sigma$ where $s_i$ is a cell, 
and~$\sigma_i$ is an action. Such a triple assigns action~$\sigma_i$ to every 
cell $s \subseteq s_i$; since a cell $s$ may be contained in many~$s_i$, we take the
triple with minimal value of $\rank_i$. Formally, given the current 
knowledge~$s$ of Player~$1$, let $(s_i,\rank_i,\sigma_i)$ 
be a triple with minimal rank in $\Pi$ such that $s \subseteq s_i$ (such a 
triple exists if~$s$ is a winning cell); the strategy represented by~$\Pi$ 
plays the action~$\sigma_i$ in~$s$.

Our implementation applies the following rules to simplify the strategies
and obtain a compact representation of winning strategies in parity games 
with imperfect information.

\medskip
\noindent(\emph{Rule~$1$}) In a strategy $\Pi$, retain only elements that
are maximal with respect to the following order:
$(s,\rank,\sigma) \succeq (s',\rank',\sigma')$
if $\rank \leq \rank'$ and $s' \subseteq s$.
Intuitively, the rule specifies that we can delete $(s',\rank',\sigma')$ 
whenever all cells contained in $s'$ are also contained in $s$; 
since $\rank \leq \rank'$, the strategy can always choose $(s,\rank,\sigma)$ 
and play $\sigma$.

\medskip
\noindent{(\emph{Rule~$2$}) In a strategy $\Pi$, 
delete all triples $(s_i,\rank_i,\sigma_i)$
such that there exists 
$(s_j,\rank_j,\sigma_j) \in \Pi$ ($i\neq j$) with $\sigma_i = \sigma_j$,
$s_i \subseteq s_j$ (and hence $\rank_i < \rank_j$ by Rule~$1$), such that
for all $(s_k,\rank_k,\sigma_k) \in \Pi$, if $\rank_i \leq \rank_k < \rank_j$ and
$s_i \cap s_k \neq \emptyset$, then $\sigma_i = \sigma_k$.
Intuitively, the rule specifies that we can delete
$(s_i,\rank_i,\sigma_i)$ 
whenever all cells contained in $s_i$ are also contained in $s_j$, 
and the action $\sigma_j$ is the same as the action $\sigma_i$. 
Moreover, if a cell $s \subseteq s_i$ is also 
contained in $s_k$ with $\rank_i \leq \rank_k < \rank_j$, then the action 
played by the strategy is also $\sigma_k = \sigma_i = \sigma_j$.


\section{\label{sec:implementation}Implementation}


Computing $\CPre(\cdot)$ is likely to require time exponential in the number of 
observations (a natural decision problem involving $\CPre(\cdot)$ is 
NP-hard~\cite{BCDHR08}). 
Therefore, it is natural to let the BDD machinery evaluate the universal 
quantification over observations in~\eqref{eq:cpre}. 
We present a BDD-based algorithm to compute $\CPre(\cdot)$.

Let $L = \{\l_1,\dots,\l_n\}$ be the state space of the game $G$.
A cell $s \subseteq L$ can be represented by a valuation $v$ of the
boolean variables $\bar{x} = x_1,\dots, x_n$ such that 
$\l_i \in s$ iff $v(x_i) = \true$, for all $1\leq i \leq n$. 
A BDD over $x_1,\dots, x_n$ is called a \emph{linear encoding}, 
it encodes a set of cells.
A cell $s \subseteq L$ can also be represented by a BDD over boolean
variables $\bar{y} = y_1,\dots, y_m$ with $m = \arup{\log_2 n}$. This is
called  a \emph{logarithmic encoding}, it encodes a single cell.

We represent the transition relation of $G$ 
by the $n \cdot \abs{\Sigma}$ BDDs 
$T_{\sigma}(\l_i)$ ($\sigma \in \Sigma$, $1\leq i \leq n$) 
with logarithmic encoding over $\bar{y}$. 
So, $T_{\sigma}(\l_i)$ represents the set 
$\{\l_j \mid (\l_i,\sigma,\l_j) \in \trans \}$.
The observations $\Obs = \{o_1,\dots,o_p\}$ are encoded 
by $\arup{\log_2 p}$ boolean variables $b_0,b_1, \dots$ 
in the BDD $B_{\Obs}$ defined by 
$$B_{\Obs} \equiv \bigwedge_{0\leq j \leq p-1} \bar{b} = [j]_2 \to  C_{j+1}(\bar{y}),$$
where $[j]_2$ is the binary encoding of $j$ 
and $C_1,\dots,C_p$ are BDDs that represent
the sets $\gamma(o_1),\dots,\gamma(o_p)$ in logarithmic encoding.

Given the antichain $q=\{s_1,\dots,s_t \}$, let $S_k$ ($1 \leq k \leq
t$) be the BDDs that encode
the set $s_k$ in logarithmic encoding over $\bar{y}$.
For each $\sigma \in \Sigma$, we compute the BDD $\CP_{\sigma}$ in
linear encoding over $\bar{x}$ as follows: 
\begin{align*}
  \CP_{\sigma} \equiv \forall \bar{b} \cdot \bigvee_{1 \leq k \leq t} \bigwedge_{1\leq i \leq n} x_i \to \big[ \forall \bar{y} \cdot 
\big(T_{\sigma}(\l_i) \land B_{\Obs}\big) \to S_k \big].
\end{align*}
Then, we define  
$\CP \equiv \bigvee_{\sigma \in \Sigma} \CP_{\sigma}(q)$,
and we extract the maximal elements in $\CP(\bar{x})$ 
as follows, with $\omega$ 
a BDD that encodes the relation of (strict) set inclusion $\subset$:
\begin{align*}
\omega(\bar{x},\bar{x}') & \equiv \Big(\bigwedge_{i=1}^{n} x_i \to x'_i\Big) \land \Big(\bigvee_{i=1}^{n} x_i \neq x'_i \Big), \\[+6pt]
\CP^{\min}(\bar{x})       & \equiv \CP(\bar{x}) \land \lnot \exists \bar{x}' \cdot \omega(\bar{x},\bar{x}') \land \CP(\bar{x}').
\end{align*}
Finally, we construct the antichain $\CPre(q)$ as the following set of BDDs in logarithmic encoding:
$\CPre(q) = \{s \mid \exists v\in \CP^{\min} : s = \{ \l_i \mid v(x_i) = \true\}  \}$.

\paragraph{Features of the tool.}
The input of the tool is a file describing the transitions and observations of 
the game graph. The output is the set of maximal winning cells, and a winning
strategy in compact representation. We have also implemented a simulator to
let the user play against the strategy computed by the tool. The user has to
provide an observation in each round (or may let the tool choose one randomly).
The web page of the tool is \verb+http://www.antichains.be/alpaga+. We provide
the source code, the executable, an online demo, and several examples.
Details of the tool features and usage are given in the appendix.

\bibliography{main}

\begin{thebibliography}{1}

\bibitem{BCDHR08}
D.~Berwanger, K.~Chatterjee, L.~Doyen, T.~A. Henzinger, and S.~Raje.
\newblock Strategy construction for parity games with imperfect information.
\newblock Technical Report MTC-REPORT-2008-005,
  http://infoscience.epfl.ch/record/125011, EPFL, 2008.

\bibitem{CDHR07}
K.~Chatterjee, L.~Doyen, T.~A. Henzinger, and J.-F. Raskin.
\newblock Algorithms for omega-regular games of incomplete information.
\newblock {\em Logical Methods in Computer Science}, 3(3:4), 2007.

\bibitem{ChaHen07}
K.~Chatterjee and T.~A. Henzinger.
\newblock Assume-guarantee synthesis.
\newblock In {\em Proc. of TACAS}, LNCS 4424, pages 261--275. Springer, 2007.

\bibitem{McNaughton93}
R.~McNaughton.
\newblock {I}nfinite games played on finite graphs.
\newblock {\em Annals of Pure and Applied Logic}, 65(2):149--184, 1993.

\bibitem{Cudd}
Fabio Somenzi.
\newblock Cudd: Cu decision diagram package.
\newblock http://vlsi.colorado.edu/\~{}fabio/CUDD/.

\bibitem{Zielonka98}
W.~Zielonka.
\newblock {I}nfinite games on finitely coloured graphs with applications to
  automata on infinite trees.
\newblock {\em Theoretical Computer Science}, 200:135--183, 1998.

\end{thebibliography}
\bibliographystyle{plain}

\newpage

\section*{Details of Tool Features}

\section{Practical implementation} 
In this section we describe the implementation details of the tool Alpaga.
\subsection{Programming Language}
Alpaga is written in Python, except for the BDD package which is
written in C. We use the CUDD BDD library~\cite{Cudd}, with its PYCUDD Python
binding. There is some performance overhead in using Python, but we
chose it for enhanced readability and to make the code easy to
change. We believe this is important in the context of
academic research, as we expect other researchers to experiment
with the tool, tweak the existing algorithms and add their own.

Alpaga is available for download at  \verb+http://www.antichains.be/alpaga+
for Linux stations. For convenience, the tool can also be tested
through a web interface (see \figurename~\ref{fig:webshot} for a
glimpse to this interface).

\begin{figure}[!tb]
 \includegraphics[width=120mm,height=120mm]{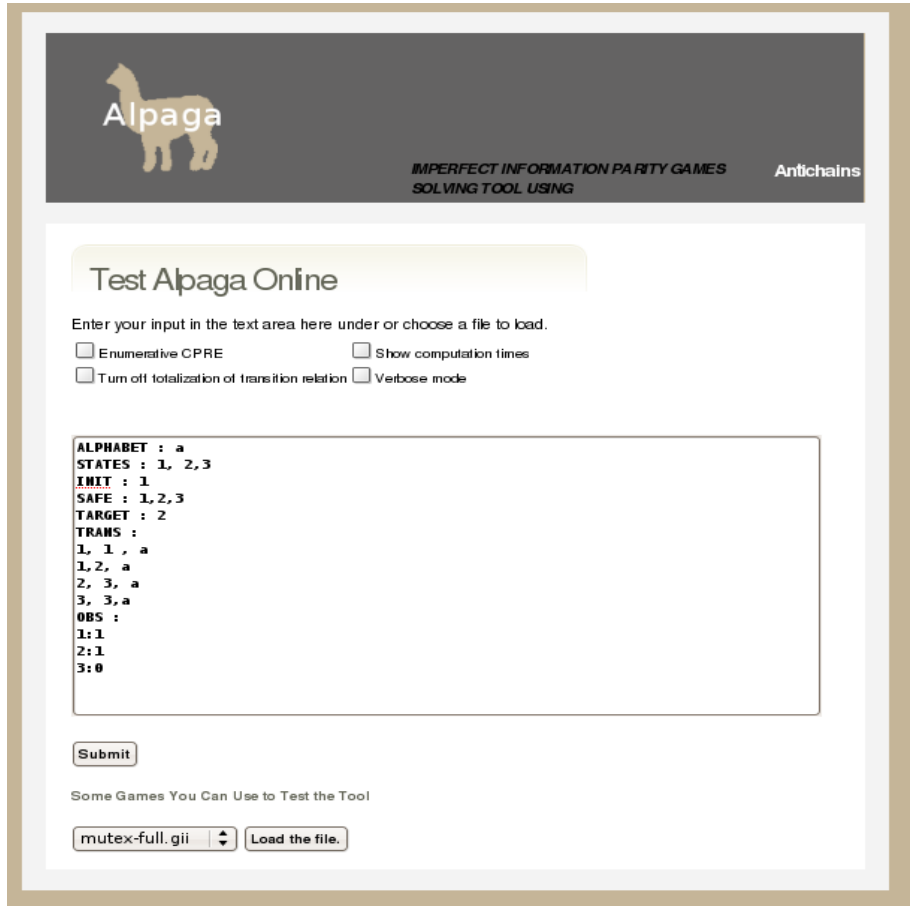}
\caption{Alpaga web interface. \label{fig:webshot}}
\end{figure}

\subsection{Code architecture}

The code consists of four main classes:
\begin{enumerate}
\item \verb|Game| is the main class of Alpaga. It encompasses all
  necessary information describing a game: BDDs for initial sets,
  target sets, observations, transition relations. 
  The class offers two implementations of the controllable predecessors operator: 
  (a)~the``enumerative'' $\CPre$ implementation which closely follows the definition of
  the $\CPre$ operator (enumerating labels, states and sets of the
  antichains, computing desired antichain intersections and unions as
  it progresses) and (b)~the $\CPre$ implementation following the BDD technique 
  explained in Section~\ref{sec:implementation}.

  Furthermore, the class offers a large set of utility functions to
  compute, for example, the successors of a set of states, its
  controllable predecessors, and to manipulate antichains of sets of
  states of the game. At a higher level, the class offers methods
  to compute strategies for specific kinds of objectives
  (ReachAndSafe: solving conjunction of reachability and safety objectives, and 
  ReachOrSafe:  solving disjunction of reachability and safety objectives).
  Finally it includes the implementation of the algorithm of~\cite{BCDHR08} 
  using all previous functions.

\item \verb|Parser|  produces an instance of the class
  \verb|Game| from an input file. The parser also offers a good amount
  of consistency checking (it checks, for example, that every state belongs
  to one and only one observation).

\item \verb|Strategy| is the class with data structure for strategy 
  representation. The description of a strategy is based on the notion
  of \emph{rank} (similar to rank of $\mu$-calculus formulas), and 
  a strategy maps a cell with a rank to a label and a cell with smaller rank.

\item \verb|StrategyPlayer| is the class implementing the interactive
  mode of Alpaga. It takes as argument a game and a strategy and
  allows the user of Alpaga to replay the strategy interactively (see below).
\end{enumerate}

\subsection{User Manual}

In this section we describe the syntax of the input file, how to read the output,
the various options of the tool, and finally we describe the interactive use 
of the tool.

\paragraph{Input.}
The syntax of the tool is straightforward and follows the formal 
description of imperfect information parity games as described in 
Section~\ref{sec-games-algo}.
Our algorithm solves games with objectives that are of the following 
form: parity objectives in conjunction with a safety objective, along 
with the disjunction with a reachability objective. 
The parity objective can be obtained as a special case when the safe set is 
the full set of states, and the target set for reachability objective is empty.
In the description below, we have the safe and target set for the safety and 
reachability objectives, respectively.

We present the following example:
\begin{figure}[!t]
    \begin{picture}(0,0)(-45,35)

\gasset{Nw=9,Nh=9,Nmr=4.5,rdist=1, loopdiam=6}


\node[Nmarks=i,NLangle=0.0](n0)(12,6){$\l_1$}
\node[Nmarks=r](n1)(32,6){$\l_2$}
\node[Nmarks=n](n2)(52,6){$\l_3$}

\node[Nmarks=n, Nw=13, Nh=13, Nmr=3, dash={1.5}0, ExtNL=y, NLangle=270, NLdist=2](A1)(12,6){$p=1$}
\node[Nmarks=n, Nw=13, Nh=13, Nmr=3, dash={1.5}0, ExtNL=y, NLangle=270, NLdist=2](A2)(32,6){$p=1$}
\node[Nmarks=n, Nw=13, Nh=13, Nmr=3, dash={1.5}0, ExtNL=y, NLangle=270, NLdist=2](A3)(52,6){$p=0$}

\drawloop[ELside=l,loopCW=y](n0){$a$}
\drawloop[ELside=l,loopCW=y](n2){$a$}


\drawedge[ELpos=50, ELside=l](n0,n1){$a$}
\drawedge[ELpos=50, ELside=l](n1,n2){$a$}

\end{picture}
\end{figure}

\begin{minipage}{7cm}
\small 
\begin{verbatim}
ALPHABET : a 
STATES : 1, 2,3
INIT : 1
SAFE : 1,2,3
TARGET : 2
TRANS : 
1, 1 , a
1,2, a
2, 3, a
3, 3,a
OBS :
1:1
2:1
3:0
\end{verbatim}
\end{minipage}

\begin{figure}[!p]
 \includegraphics[width=120mm]{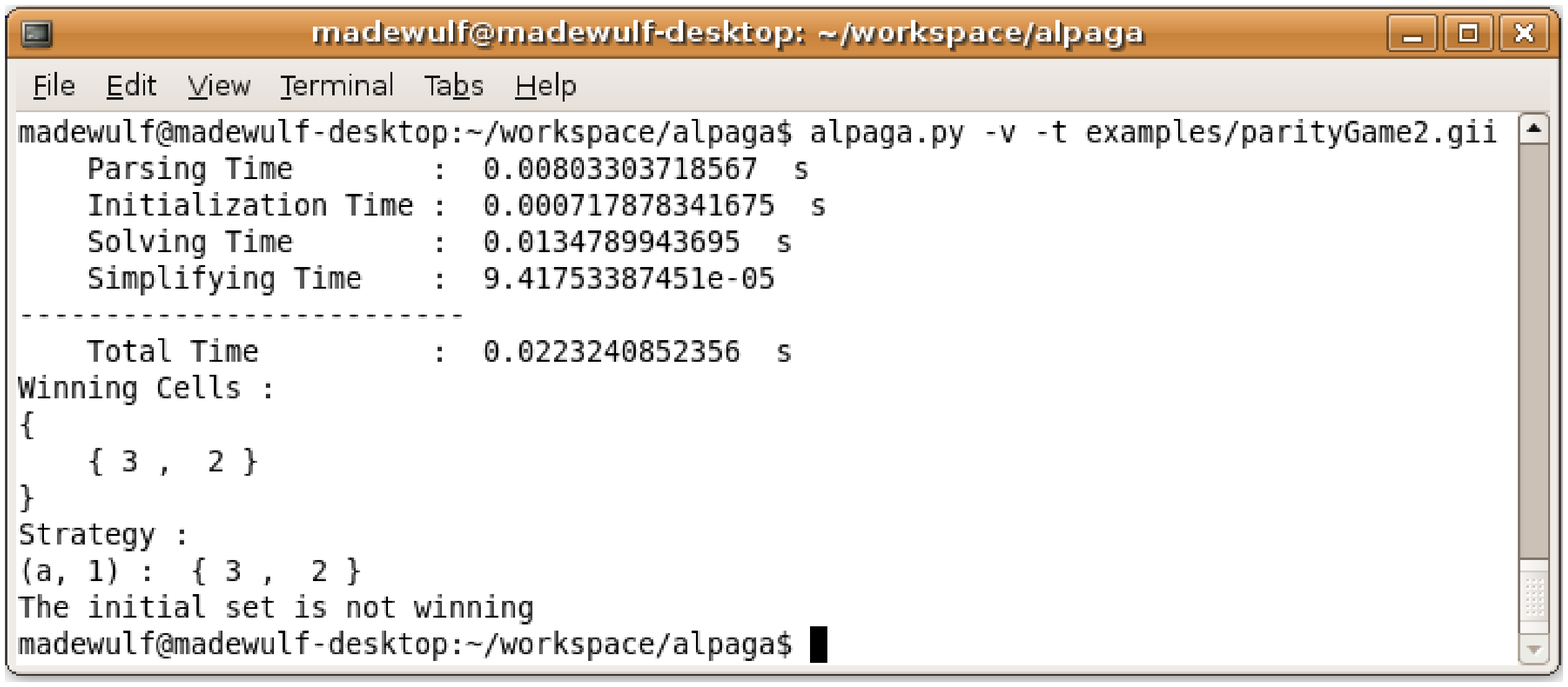}
\caption{Output of Alpaga.\label{fig:game2}}
\end{figure}

\bigskip

The input file describing a parity game with imperfect information is
constructed as follows:
\begin{itemize}
\item the sets of labels, states, initial states, safe states, and
  target states are all specified on a single line introduced by the
  corresponding keyword ALPHABET, STATES, INIT, SAFE or TARGET. The
  name of the states and labels can be any string accepted by Python
  that does not include a blank space or the \# character (which is 
  used for comments). However, the keyword SINK is reserved (see below).
\item the transition relation is defined on a sequence of lines
  introduced by the keyword TRANS on a single line. After the
  TRANS keyword, each line specifies on a single line a transition, by
  giving the initial state, the destination state and the label of the
  transition, all separated by commas.
\item Finally, the observations and corresponding priorities are
  specified in a similar fashion. They are introduced by the keyword
  OBS on a single line. Then follows the specification of the
  observations. Each observation is specified on a single line as a
  set of comma-separated states, followed by its priority (a positive
  integer number) which must be preceded by a colon.
\item Blank lines are allowed anywhere as empty comments. Nonempty 
  comments start with 
  the character \# and extend to the end of the line.
\end{itemize}

\paragraph{Output.} The tool output for our example is  
in \figurename~\ref{fig:game2}.
The winning strategy computed by the tool is represented by a list of
triples $(\sigma,\rank,s) \in \Sigma \times \nat \times 2^L$ where $\sigma$ is 
an action, and $s$ is a cell. 
The strategy is represented in the compact form after applying Rule~1 and 
Rule~2 for simplification of strategies.
The strategy representation can be used to find the action to play, given the 
current knowledge $s'$ of Player~$1$ as follows: play the action $\sigma$ such 
that $(\sigma,\rank,s)$ is a triple in the list with minimal rank such that 
$s' \subseteq s$ (such a triple must exist if $s'$ is a winning cell).

\paragraph{Tool options.}

We now describe the various options with which the tool can be used.
\begin{verbatim}
alpaga.py [options] file
\end{verbatim}
The possible options are the following:
\begin{itemize}
\item -h Shows an help message and exits.
\item -i After computing a strategy, launches the interactive strategy
  player which allows to see how the strategy computed by the tools
  executes in the game. In this mode, the tool shows which move
  is played by the strategy, given the current knowledge (i.e., a set of states 
  in which the player can be sure that the game is - the initial knowledge 
  is the set of initial states). Then the tool allows to choose the
  next observations among the observations that are compatible with the
  current knowledge.
\item -e Uses the enumerative $\CPre$ in all computations. There are two
  different implementations for the controllable predecessor operator
  ($\CPre$), one temporarily using a linear encoding of the
  resulting antichain for the time of the computation, and an
  enumerative algorithm following closely the definition of the $\CPre$
  operator.
\item -n Turns off the totalization of the transition relation. By
  default, Alpaga completes the transition relation so that it becomes
  total, which means that a transition of every label exists from
  each state. Therefore, Alpaga first adds a state named SINK with
  priority 1 (corresponding to a new observation), from which every label loops
  back to SINK, and then adds a transition
\begin{verbatim}
      s, SINK, lab
\end{verbatim}
  for each pair (s, lab)  such that there does not exist a transition
  from state s on label lab. Note that the name SINK is reserved.
\item -r Turns on the display of stack traces in case of error.
\item -s Turns off the simplification of the strategies before display.
\item -t Displays computation times, which includes time for parsing
  the file (and constructing the initial BDDs), time for
  initializing the linear encoding, for computing a strategy, and
  for simplifying that strategy.
\item -v Turns on the display of warnings, which mainly list the
  transitions added by the totalization procedure.
\end{itemize}

\paragraph{Interactive mode.}
After computing a strategy for a parity game, the tool can switch to
interactive mode, where the user can ``replay'' the strategy, to check
that the modelization was correct. 
The user of Alpaga plays the role of Player~$2$,
choosing the observation among the compatible observations
available, and getting the resulting knowledge of player 1
and which move she will play.

Practically, in interactive mode, type help for the list of commands:
the standard way for playing a strategy is the following: launch
alpaga with option -i, type go at the interactive prompt, type the
number of an observation, type enter twice,
repeat. \figurename~\ref{fig:interactive} shows an interactive
Alpaga session.

\begin{figure}[!tp]
 \includegraphics[width=120mm]{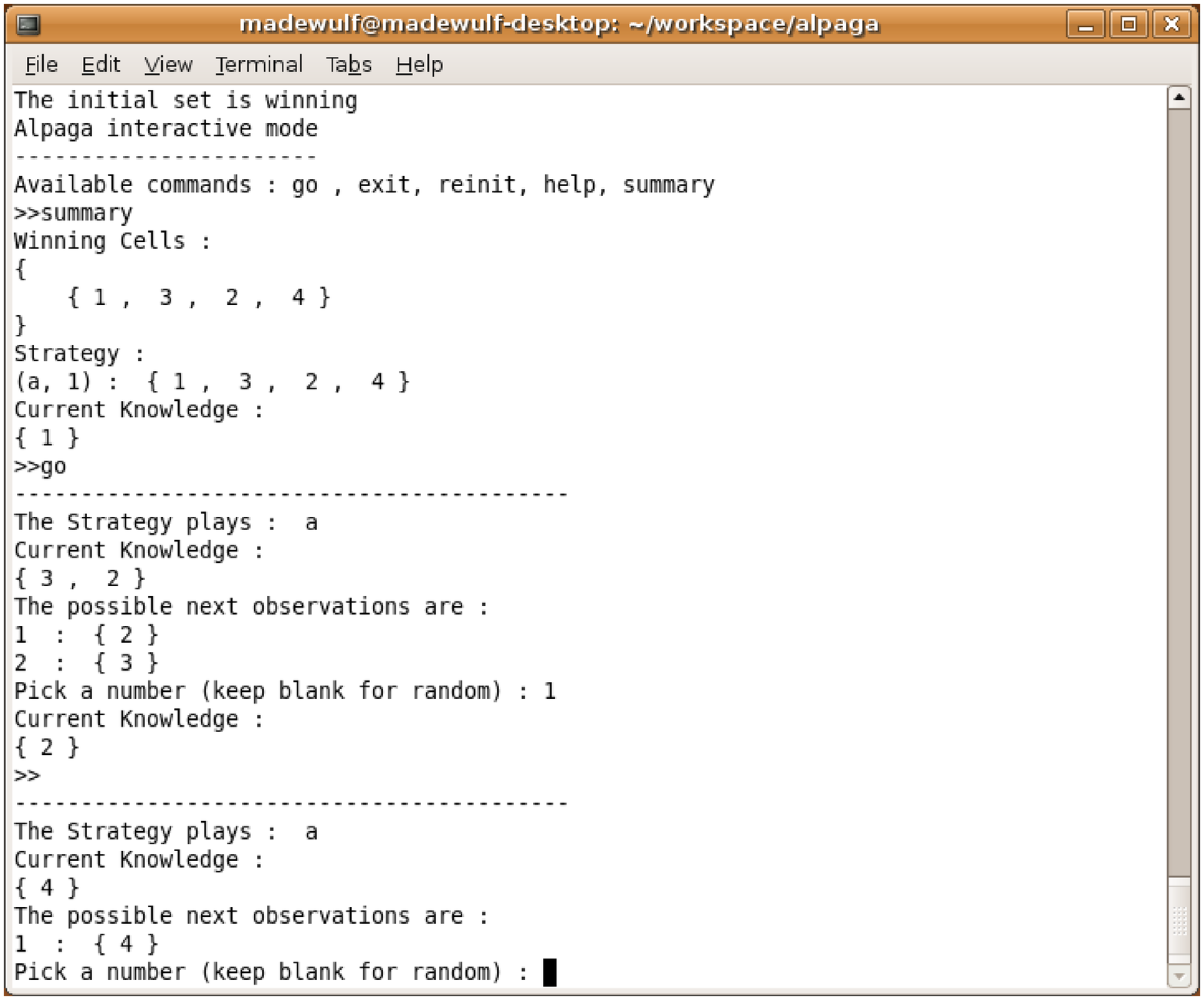}
\caption{\label{fig:interactive}Interactive strategy player of Alpaga.}
\end{figure}

\section{Example: mutual-exclusion protocol}

We demonstrate the use of games with imperfect information
to synthesize reactive programs in distributed systems.
%
%
We consider the design of a mutual-exclusion protocol for two processes, 
following the lines of~\cite{ChaHen07}. We assume that one process (on the 
right in \figurename~\ref{fig:mutex1}) is completely specified. The second
process (on the left in \figurename~\ref{fig:mutex1}) has freedom of choice
in line~$4$. It can use one of~$8$ possible conditions~$\mathtt{C1}$--$\mathtt{C8}$ 
to guard the entry to its critical section in line~$5$. The boolean variables 
$\mathtt{flag[1]}$ and $\mathtt{flag[2]}$ are used to place
a request to enter the critical section. They are both visible to each process.
The variable $\mathtt{turn}$ is visible and can be written by the two processes. 
Thus, all variables are visible to the left process, except the program counter
of the right process.

There is also some nondeterminism in the length of the delays in lines~$1$ and~$5$ of the two 
processes. The processes are free to request or not the critical section 
and thus may wait for an arbitrary amount of time in line~$1$
(as indicated by $\mathtt{unbounded\_wait}$), but they have to leave the
critical section within a finite amount of time (as indicated by $\mathtt{fin\_wait}$).
In the game model, the length of the delay is chosen by the adversary.

Finally, each computation step is assigned to one of the two processes by a 
\emph{scheduler}. We require that the scheduler is fair, i.e. it assigns
computation steps to both processes infinitely often. In our game model, we encode
all fair schedulers by allowing each process to execute an arbitrary finite
number of steps, before releasing the turn to the other process. Again, the 
actual number of computation steps assigned to each process is chosen
by the adversary.


The mutual exclusion requirement (that the processes are never simultaneously
in their critical section) and the starvation freedom requirement
(that whenever the left process requests to enter the  critical section, then it 
will eventually enter it) can be encoded using three priorities.

When solving this game with our tool, we find that Player~$1$ is winning, 
and that choosing $C_8$ is a winning strategy.

\begin{figure}[!tb]
\begin{center}
\begin{minipage}{7cm}
\small
\begin{verbatim}
do {
unbounded_wait;
flag[1]:=true;
turn:=2;

| while(flag[1]) nop;           (C1)
| while(flag[2]) nop;           (C2)
| while(turn=1) nop;            (C3)
| while(turn=2) nop;            (C4)
| while(flag[1] & turn=2) nop;  (C5)
| while(flag[1] & turn=1) nop;  (C6)
| while(flag[2] & turn=1) nop;  (C7)
| while(flag[2] & turn=2) nop;  (C8)

fin_wait;  // Critical section
flag[1]:=false;
} while(true)
\end{verbatim}
\end{minipage}%
\begin{minipage}{5.5cm}
\begin{verbatim}
do {
unbounded_wait;
flag[2]:=true;
turn:=1;

while(flag[1] & turn=1) nop;








fin_wait;  // Critical section
flag[2]:=false;
} while(true)
\end{verbatim}
\end{minipage}%
\end{center}
\caption{Mutual-exclusion protocol synthesis.}
\label{fig:mutex1}
\end{figure}

\end{document}